 \documentclass[final,5p,times,twocolumn,authoryear]{elsarticle}

\usepackage{amssymb}
\usepackage{lipsum}
\usepackage{amsmath}
\usepackage{graphicx}
\newcommand{\kms}{km\,s$^{-1}$}

\journal{Annals of Physics}

\begin{document}

\begin{frontmatter}



\title{Quasi-superfluid and Quasi-Mott phases of strongly interacting bosons in shallow optical lattice}


\author[first]{Subhrajyoti Roy}

\author[second]{Rhombik Roy \fnref{haifa}}
\fntext[haifa]{Current affiliation: Department of Physics, University of Haifa, 3498838 Haifa, Israel.}

\author[third]{Arnaldo Gammal}

\author[second,third]{Barnali Chakrabarti \corref{cor1}}
\cortext[cor1]{email: barnali.physics@presiuniv.ac.in}

\author[fifth]{Budhaditya  Chatterjee}

\affiliation[first]{organization={Indian Statistical Institute},%Department and Organization
            %addressline={}, 
            city={Tezpur},
            postcode={784501}, 
            state={Assam},
            country={India}}

\affiliation[second]{organization={Department of Physics, Presidency University},%Department and Organization
            %addressline={ }, 
            city={86/1 College Street, Kolkata},
            postcode={700073}, 
            state={West Bengal},
            country={India}}

\affiliation[third]{organization={Instituto de Física, Universidade de Sao Paulo},%Department and Organization
            %addressline={Indian Statistical Institute}, 
            city={Sao Paulo},
            postcode={05508-090}, 
            %state={Assam},
            country={Brazil}}

%\affiliation[fourth]{organization={Instituto de Física, Universidade de Sao Paulo},%Department and Organization
            %addressline={Indian Statistical Institute}, 
%            city={Sao Paulo},
%            postcode={05508-090}, 
            %state={Assam},
%            country={Brazil}}

\affiliation[fifth]{organization={Department of Physics, Chandigarh University},%Department and Organization
            %addressline={}, 
            city={Mohali},
            postcode={140413}, 
            state={Punjab},
            country={India}}

\begin{abstract}
We explore the ground states of strongly interacting bosons in the vanishingly small and weak lattices using the multiconfiguration time-dependent Hartree method for bosons (MCTDHB) which calculate numerically exact many-body wave function. Two new many-body phases: fragmented or quasi superfluid (QSF) and incomplete fragmented Mott or quasi Mott insulator (QMI) are emerged due to the strong interplay between interaction and lattice depth.  Fragmentation is utilized as a figure of merit to distinguish these two new phases. We utilize the eigenvalues of the reduced one-body density matrix and define an order parameter that characterizes the pathway from a very weak lattice to a deep lattice. We provide a detailed investigation through the measures of one- and two-body correlations and information entropy. We find that the structures in one- and two-body coherence are good markers to understand the gradual built-up of intra-well correlation and decay of inter-well correlation with increase in lattice depth.
\end{abstract}



\begin{keyword}
%% keywords here, in the form: keyword \sep keyword, up to a maximum of 6 keywords
Phases in optical lattice \sep Many-body physics \sep Shallow lattice \sep MCTDHB



\end{keyword}

\end{frontmatter}

%\tableofcontents



\section{Introduction}\label{intro}
Ultracold atomic gases provide unprecedented experimental control to study complex many-body systems in lattice~\cite{RevModPhys.80.885,RevModPhys.83.863,rr1}. Experimentally it is quite straightforward to tune the lattice depth of the periodic optical potential and subsequent study of different phases of ultracold atoms~\cite{dimension3,nat.phys.8.325,nature.415,nature.419} The interplay between the lattice depth and interatomic interaction triggers the superfluid-to-insulator transition commonly known as Mott transition~\cite{nature.415,science.1192368,PhysRevLett.81.3108}. The Mott transition takes place in deep lattice~which is characterized by a high localization of the atoms and theoretically the Bose-Hubbard (BH) model essentially captures the underlying process~\cite{book1}. Later, many different extensions of the standard BH model have been introduced, specially when interatomic interactions are strong enough which leads to interesting phenomenon~\cite{Rep}. Nevertheless, there are open questions on the validity of BH model for ultracold gases in shallow lattice. In many aspects, the experiments with ultracold atoms in shallow lattice are easily realized. In Ref. ~\cite{Haller:2010,haller}, amplitude of the periodic optical potential is varied from zero to large value and the Mott transition in the shallow lattice has been observed by modulation spectroscopy and transport measurements. The pinning transition in the vanishing lattice is described by (1+1) quantum sine-Gordon model~\cite{Coleman}. Later, complete phase diagram of the continuous model is mapped out~\cite{Buchler,Haque,Gregori,Boeris} and it is demonstrated that the sine-Gordon model used for shallow lattice is insufficient. \\

In the present work, we address the system implemented in the experiment~\cite{Haller:2010}, one dimensional strongly interacting bosons in optical lattice of variable strength. The main purpose of the work is to understand the few bosons system utilizing numerically exact solution of Schr\"odinger equation and to explain many-body phases. The study of strong quantum correlation in a smaller system with a very finite number of interacting particles are receiving special interest in the recent experiments~\cite{Nature385,Nature606}. We use the multiconfigurational time-dependent Hartree method for bosons (MCTDHB)~\cite{Streltsov:2006,Streltsov:2007,Alon:2007,Alon:2008,Lode:2016,Fasshauer:2016,Lode:2020}, implemented in the MCTDH-X software~\cite{Lin:2020,MCTDHX}. We observe strong competition between the dimensional confinement and quantum correlation. We find the emergent phases for strongly interacting bosons in extremely weak lattice are neither the ideal superfluid (SF) nor the ideal Mott (MI) phases. In the limit of shallow potential, when the lattice effect is subrelevent, many-body phases are organized and controlled by the quantum correlation originated from strong interatomic interaction. The many-body physics of such strongly correlated system lies beyond the scope of BH model. We utilize 'fragmentation' as the key measure to distinguish the emergent phases. We quantify ``excited fragmentation" as the fraction of atoms contributed by the other natural orbitals except the one with the largest occupation. Fragmentation is the hallmark of MCTDHB which utilizes multiorbitals ansatz and the strongly interacting bosons may be fragmented even in a very weak or vanishingly small lattice. \\

The present work provides a complete path for varying lattice depth $V_0$ = $0.1E_r$ to $8.0E_r$, for fixed inter-atomic interaction which corresponds to strongly interacting repulsive bosons. The system consists of three strongly interacting bosons in three wells. We calculate the many-body wave function with very high precision, and one-body and two-body correlations are calculated accurately-thus our elemental building block is able to capture all essential many-body physics in the larger lattice. In our small ensemble, the term 'quantum phase' is not rigorously applicable. However, the calculated many-body states can be taken as few-body precursor of thermodynamic phases. For clarity we term the few-body states as 'phase'. The two key measures : fragmentation and correlation are utilized to describe the different phases in varying lattice depth. For almost vanishing lattice ($V_0$ $\simeq$ $0.5E_r$); several natural orbitals have significant population; thus the many-body phase is fragmented.  It also exhibits strong diagonal correlation and weak off-diagonal correlation in one-body and clear signature of anti-bunching effect in two-body correlations. 
It is not a SF phase which is designated by macroscopic occupation in one single orbital and uniform global correlation across the lattice. It is also not a Mott phase as it does not exhibit complete localization in each well. Quantitative measures as discussed in the result section establish that the many-body phase is best diagnosed as fragmented superfluid or quasi SF (QSF) phase. For weak lattice ($V_0$ $\simeq$ $2.0E_r$); the system becomes more fragmented but does not exhibit $S$-fold fragmentation, where $S$ is the number of lattice sites. The diagonal correlation in one-body becomes stronger at the cost of weaker off-diagonal correlation. So the many-body phase is close to Mott but not a perfect Mott. The quantitative estimates establish that the many-body phase can be designated as incomplete fragmented Mott or quasi-Mott (QMI) phase. For larger lattice depth ($V_0$ $\simeq$ $5.0E_r$), completely fragmented Mott phase is achieved. \\

All these phases are studied by key quantities: a) occupation in natural orbital, b) excited fragmentation, c) order parameter d) one-body and two-body correlations and e) information entropy measures. All these measures are synchronized and clearly define the many-body phases in vanishing to weak to deep lattices.  \\  

The paper is organized as follows. In Sec. II, we introduce the theoretical framework. Sec III introduces the quantities of interest. In Sec IV, we present the results. Sec V concludes our observations.

\section{System and method}\label{numerics}
For $N$ bosons in a quasi-one-dimensional optical lattice, the Hamiltonian is 
\begin{equation} 
\hat{H}= \sum_{i=1}^{N} \hat{h}(x_i) + \lambda_0\sum_{i<j=1}^{N}\delta(x_i-x_j)
\end{equation}
We rescaled the Hamiltonian by the lattice recoil energy $E_r = \frac{\hbar^{2}k^{2}}{2m}$ to achieve the convenient dimensionless units; $\hbar$=$m$=$k$= $1$.
 $\hat{h}(x) = \hat{T}(x) + \hat{V}_{trap}(x)$ is the one-body Hamiltonian. $\hat{T}(x)$ is the kinetic energy operator; $\hat{T}(x) = -\frac{1}{2} \sum_{i=1}^{N} \frac{\partial^2}{\partial x^2}$ and $\hat{V}_{trap}(x)$ is the external trapping potential. $V_{trap}(x)$ = $V_{0}\sin^{2}(kx)$. We consider three lattice sites with periodic boundary conditions, the strong transverse confinement ensures a quasi-1D trap. We choose strong repulsive interaction of strength $\lambda_{0}=10.0$. \\
 The MCTDHB method represents a bosonic variant within the broader family of MCTDH methodologies~\cite{10.1063/1.4821350,10.1063/1.2902982,BECK20001,10.1063/1.1580111,HALDAR201872,Wang2015,Bhowmik2022,PhysRevA.91.012509,variational4,Lévêque_2017} in which, the many-body wavefunction is constructed as a linear combinations of the time dependent permanents and time-dependent complex-valued coefficients,
 \begin{equation}
\vert \psi(t)\rangle = \sum_{\vec{n}}^{} C_{\vec{n}}(t)\vert \vec{n};t\rangle,
\label{many_body_wf}
\end{equation}
where
\begin{equation}
\vert{\vec{n};t} \rangle=\vert{n_1,n_2,\dots,n_M;t}\rangle=\prod_{i=1}^M\left[\frac{(b_i^\dagger (t))^{n_i}}{\sqrt{n_i!}} \right]\vert{\text{vac}}\rangle. 
\end{equation}
Each operator $b_k^\dagger(t)$ 
creates a boson occupying the time-dependent single-particle state (orbital) $\phi_k(x,t)$.
Note that the expansion coefficients $\left \{C_{\vec{n}}(t)\right\}$ as well as  orbitals $\left\{ \phi_i (x,t)\right\}_{i=1}^{M}$ that build up the permanents $\vert \vec{n};t\rangle$ are explicitly time dependent and fully variationally optimised quantities~\cite{Streltsov:2007,Alon:2008}. For $N$ bosons distributed over $M$ orbitals, the number of permanents become $N_{conf}$=  $ \left(\begin{array}{c} N+M-1 \\ N \end{array}\right)$. Thus, in the limit of $M \rightarrow \infty$, the wave function becomes exact, and the set $ \vert n_1,n_2, \dots,n_M \rangle$ spans the complete $N$-particle Hilbert space. For practical calculation, we restrict the number of orbitals to the desired value requiring the proper convergence in the measured quantities. 
Compared to a time-independent basis, as the permanents are time-dependent, a given degree of accuracy is reached with much shorter expansion~\cite{mctdhb_exp1,mctdhb_exp2}. We also emphasize that MCTDHB is more accurate than exact diagonalization which uses the finite basis and is not optimized. Whereas in MCTDHB, as we use a time adaptive many-body basis set, it can dynamically follow the building correlation due to inter-particle interaction~\cite{Alon:2008,Alon:2007, mctdhb_exact3,barnali_axel} and it has been widely used in different theoretical calculations~\cite{rhombik_jpb,rhombik_pra,rhombik_quantumreports, rhombik_epjd,rhombik_aipconference,rhombik_pre,rhombik_scipost}. To get the ground state, we propagate the MCTDHB equations in imaginary time.
As in the shallow lattice, the many-body wavefunction shows strong fragmentation--we discuss the issue of convergence of simulation with MCTDHB. The convergence is ascertained in the following ways. We systematically increase the number of orbitals until the basic measures like occupation in natural orbitals and degree of fragmentation converge. Additionally, the convergence is further confirmed when the occupation in the highest orbital is negligible. 

 \section{Quantity of Interest}
\subsection{Natural occupation and excited fragmentation}
Natural occupation and the excited fragmentation are the two key measures to understand the transition from QSF to QMI to MI with increase in lattice depth. Excited fragmentation is defined as the fraction of atoms that do not occupy the lowest eigenstate of the reduced one-body density matrix or the contribution coming from other orbitals except the one that is maximally populated. 
The reduced one-body density matrix is 
\begin{equation}
\rho^{(1)}(x,x^{\prime}) = \sum_k \rho_k^{(NO)}\phi_{k}^{*(NO)}(x) \phi_{k}^{(NO)}(x^{\prime}).
\end{equation}
$\phi_{k}^{(NO)}(x)$ are the eigenfunctions of $\rho^{(1)}$ and known as natural orbitals. $\rho_k^{(NO)}$ are the eigenvalues of $\rho^{(1)}$ and known as natural occupations. In the mean-field perspective, when one single natural orbital is occupied, $\rho^{(1)}$ has only a single macroscopic eigenvalue, and the many-body state is non-fragmented. The excited fragmentation $F$
becomes zero. However multi-orbital ansatz facilitates to study of the beyond mean field picture-several natural orbitals may exhibit significant population and $F$ may become non-zero. The corresponding state is fragmented. 
\subsection{Order parameter}
We define an order parameter 
\begin{equation}
\bigtriangleup = \sum_{i} \left(  \frac{n_i}{N} \right) ^2
\end{equation}
where $n_i$ is the natural occupation in $i^{th}$ orbital. 
For the SF phase, $\bigtriangleup=1$ and only one eigenvalue is nonzero. For the Mott phase, the number of significantly contributing orbitals becomes equal to the number of sites in the lattice, and $\bigtriangleup$ = $\frac{1}{S}$, $S$ is the number of lattice sites. Thus $\bigtriangleup=1$ and $\bigtriangleup$ = $\frac{1}{S}$ are the two extreme values for the two known phases. We find $\bigtriangleup$ is a good marker to identify the QSF and QMI phases in the shallow lattice. 
\subsection{Glauber correlation function}
Understanding a quantum many-body phase implies understanding the role of correlations. For a strongly interacting system, correlation functions of higher orders are necessary to understand the complex phases. 
The $p$-th order Glauber correlation function $g^{(p)}(x_{1}^{\prime}, \dots, x_{p}^{\prime}, x_{1}, \dots, x_{p};t)$ measures degree of spatial coherence and
correlations within the system. It is defined as:
\begin{equation}
\begin{split}
g^{(p)}(x_{1}^{\prime}, \dots, x_{p}^{\prime}, x_{1}, \dots, x_{p}) = \\ \frac{\rho^{(p)}(x_{1}, \dots, x_{p} \vert x_{1}^{\prime}, \dots, x_{p}^{\prime})}{\sqrt{\prod_{i=1}^{p} \rho^{(1)} (x_i \vert x_i) \rho^{(1)} (x_{i}^{\prime} \vert x_{i}^{\prime})}}.
\label{correlation}
\end{split}
\end{equation}
where $\rho^{(p)}(x_{1}^{\prime}, \dots, x_{p}^{\prime} \vert x_{1}, \dots, x_{p})$ is the $p$-th order reduced density matrix and describes the joint probability distribution of finding $p$ particles at specific positions and defined as
\begin{equation}
\begin{split}
\rho^{(p)}(x_{1}^{\prime}, \dots, x_{p}^{\prime} \vert x_{1}, \dots, x_{p}) = \frac{N!}{(N-p)!}\int dx_{p+1},...dx_{N} \\ \psi^{*}(x_{1}^{\prime},\dots, x_{p}^{\prime},x_{p+1},\dots,x_{N}) \psi(x_{1},\dots,x_{p},x_{p+1}\dots,x_{N}).
\label{pbodydensity}
\end{split}
\end{equation}
The diagonal elements of $g^{(p)}(x_{1}^{\prime}, \dots, x_{p}^{\prime}, x_{1}, \dots, x_{p})$, denoted as $g^{(p)}( x_{1}, \dots, x_{p})$, provide a measure of $p$-th order coherence. 
If $\vert g^{(p)}( x_{1}, \dots, x_{p};t) \vert = 1$, the system is fully coherent, while deviations from unity indicate partial coherence. Specifically, $g^{(p)}( x_{1}, \dots, x_{p};t) > 1$ implies correlated detection probabilities at positions $x_{1}, \dots, x_{p}$, while $g^{(p)}( x_{1}, \dots, x_{p};t) < 1$ indicates anti-correlations.
These measures collectively provide valuable insights into the evolving density profile,coherence and are especially useful to understand the interplay between inter-well and intra-well correlations.

\subsection{Many-Body entropy}
We employ the measures of information entropy as another marker to understand the pathway from QSF to QMI and finally to the MI phase. As in MCTDHB, both the coefficients and the orbitals contribute to the many-body wavefunction ansatz, we define two measures of many-body information entropy.  
$S^{info}$ is defined as
\begin{equation}
 S^{info} = -\sum_{\vec{n}} | C_{\vec{n}}|^2 \ln |C_{\vec{n}}|^{2}.
\end{equation}
$ S^{info}$ measures the effective number of basis states that contribute to a given many-body state. Besides $ S^{info}$ we are also interested in the calculation of occupation entropy defined as 
\begin{equation}
 S^{occu}= - \sum_{i} n_i [\ln n_i].
\end{equation}
$S^{occu}$ is an entropy obtained from the natural occupation. 
For the Gross Pitaevskii  mean-field theory, one has $ S^{info}$ =0, as one single coefficient contributes, $ S^{occu}=0 $ always as there is only one natural occupation $\vec{n}_1=\frac{n_1}{N}=1$ in this case. For multiorbital theories, several occupation numbers can be different from $0$, and the magnitude of  $S^{info}$ and $S^{occu}$ indicates how well the state could be described by a mean-field approach. In our investigation, the study of many-body entropy plays a crucial role. In the SF phase, a single orbital and coefficient suffice to describe the state, resulting in both $S^{info}$ and $S^{occu}$ displaying zero values. Conversely, in the Mott insulator phase, the system becomes $\frac{1}{S}$-fold fragmented, where $S$ denotes the number of lattice sites. Consequently, both $S^{info}$ and $S^{occu}$ exhibit non-zero values. We utilize the Gaussian Orthogonal Ensemble (GOE) theory for comparative analysis~\cite{Kota,barnali_axel,PhysRevLett.112.170601}. Thus, during the transition from the SF to MI phase, $S^{info}$ and $S^{occu}$ make transition from zero to the GOE value, facilitating the detection of intermediate phases such as quasi-superfluid (QSF) and quasi-Mott insulator (QMI). \\

\section{Results}
\subsection{Fragmentation and order parameter}
Our set up consists of $N=3$ bosons in triple well set up $S=3$ with periodic boundary condition. Interaction strength is fixed for $\lambda_0=10.0$ throughout our calculation. We fix the number of orbitals to $M=12$, when the occupation in the last orbital is less than $10^{-5}$. We have also validated the convergence of our calculations by repeating the computations with $M=15$ orbitals, consistently yielding identical results for the measured quantities. We simulate the emergence of few-body phases as a function of varying lattice depth. We demonstrate how the natural populations in different orbitals are used to construct the mesoscopic ``order parameter". The concept of order parameter is generally used in the thermodynamic limit. However, for our choice of a small ensemble, the ground state properties that are presented here are analogous to the macroscopic phases.  The few-body physics remains the same with five bosons in five wells. Thus ``order parameter" can be taken as a finite-sized precursor of the quantum phases.\\

Figures 1(a), 1(b) and 1(c) show a plot of the eigenvalues $\rho_k^{(NO)}$, excited fraction (fragmentation), and the order parameter respectively as a function of lattice depth. We plot the occupation in the lowest three natural orbitals $\left( \rho_1^{(NO)}, \rho_2^{(NO)}, \rho_3^{(NO)} \right)$ which are significantly occupied. For almost vanishing lattice ($V_0=0.1 E_r$), the lowest orbital has slightly more than $70 \%$ population, and the other two orbitals contribute slightly more than $10 \%$ (each). So the state is neither $\vert{1,0,0}\rangle$ (which corresponds to SF) nor $\vert{1,1,1}\rangle$ (which corresponds to MI) configuration. However, as the population in the first natural orbital is significantly high compared to the two other orbitals, we call it fragmented SF or QSF. With an increase in $V_0$, the occupation of the first natural orbital gradually decreases, and that in the second and third natural orbitals increases which leads to complete three-fold fragmentation for deep lattice ($V_0=8.0E_r$). In weak lattice ($V_0 \simeq 2 E_r$), the many-body state is not exactly three-fold fragmented, but the population imbalance in the lowest three orbitals is minimum, we call it the incomplete fragmented Mott phase (QMI). At deep lattice three-fold fragmentation is described by $33.3 \%$ population of each three lowest orbitals, it is the unique feature of the MI phase. Thus strongly interacting bosons in shallow lattices exhibit two new many-body phases, termed QSF and QMI. They are the {\it {hallmark}} states of MCTDHB and beyond the scope of any other existing model calculations. \\

We quantify the degree of fragmentation by introducing excited fragmentation $(F)$. $F$ determines the function of atoms that are outside the natural orbital of the highest population. In Fig. 1(b), we plot $F$ as a function of varied lattice depth. As $F > 0$ for the entire range of lattice depth, the system is always fragmented. With the increase in $V_0$, fragmentation increases, and at the deep lattice, it saturates to 0.66 which refers to a three-fold fragmented many-body state and can be configured as $\vert{1,1,1} \rangle$ configuration. To get more insight into the fragmented SF and non-fragmented Mott phase, we introduce an order parameter (Eq.(5)). In Fig. 1(c), we plot $\bigtriangleup$ for various lattice depth potentials.
The two extreme limits are $1.0$ (SF) and $\frac{1}{S}$=0.33 (MI). The order parameter does not coincide with the SF or MI phases for vanishingly small and weak lattices. 
Fig. 1(c) clearly exhibits that the order parameter scales QSF to QMI to MI phases. \\
\begin{figure}
\centering
\includegraphics[scale=0.20, angle=-0]{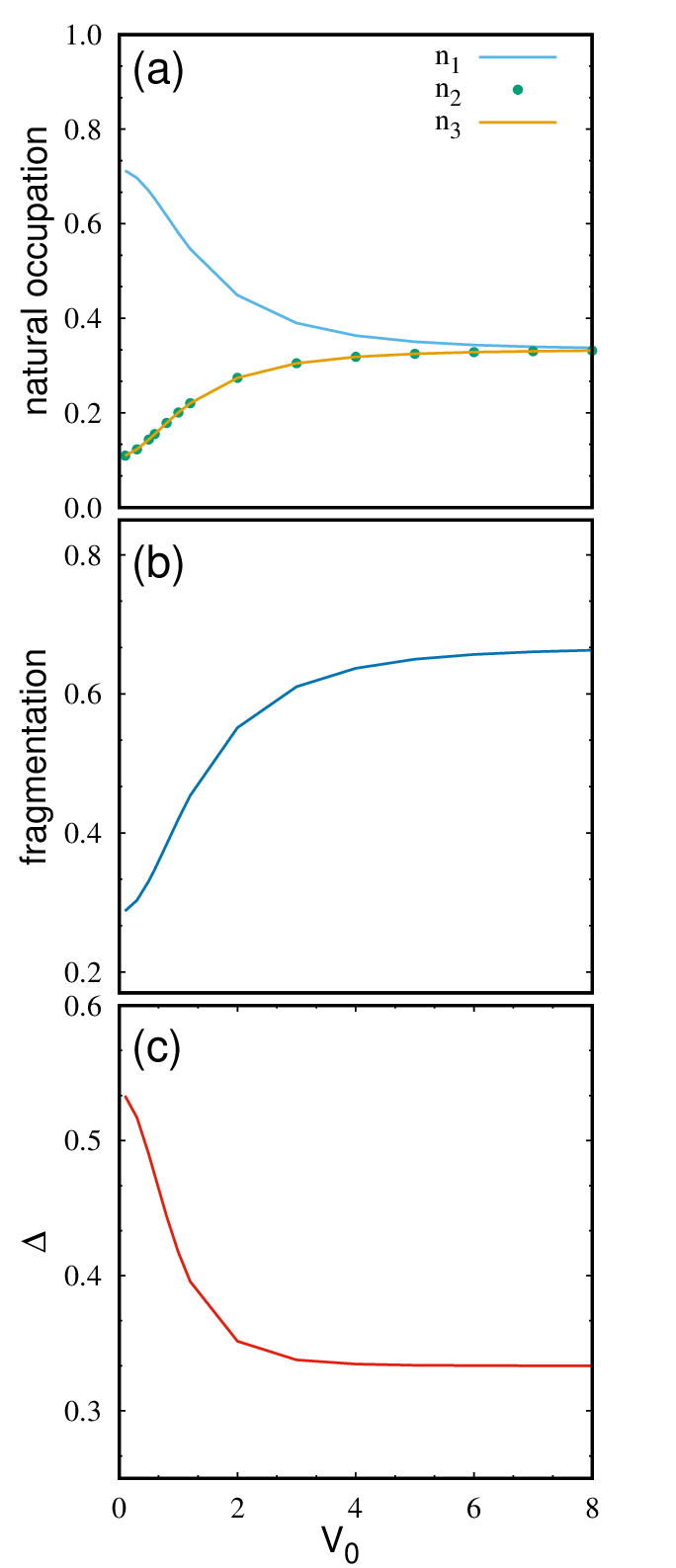}
\caption{(a) Occupations in the three lowest significantly populated orbitals, (b) Fraction of atoms outside the natural orbital having maximum occupation, (c) order parameter, all as a function of lattice depth $V_0$ and for fixed interaction strength $\lambda_0=10.0$. The emergence of fragmentation and subsequent growth in fragmentation, signify that the system enters a strongly correlated new phase. All quantities are dimensionless, see text for additional details.}

\label{fig1}
\end{figure}
\subsection{Delocalization in Fock space}
The pathway from fragmented SF to incomplete fragmented Mott state to one-third fragmented Mott state can be well understood by the ``partial delocalization to complete delocalization" in the many boson Fock space. In Fig.~2, we plot the expansion coefficients $|C_{\vec{n}}|^2$ as a function of the index of the basis states for various choices of lattice depth. Although numerical computation is done with $M=12$ orbitals, but the lowest three orbitals are significantly populated, as also seen in Fig.~1(a). With  $N=3$ bosons in $M=3$ orbitals, $N_{conf}$ = 340. So, we presented our results for index up to $340$. The index is determined based on the vector $\vec{n}$ through the mapping outlined in reference~\cite{fork}.
%With increasing time, more coefficients in the expansion become significant, but the spread is far from the whole available space spanned by the D = Nconf = 3003 configurations. The state stays rather localized throughout the quench dynamics. 
If the number of nonzero elements of $\left \{C_{\vec{n}}\right\}$ occupy a very small portion of $N_{conf}$, we refer it to a localized many-body state. Localized states closely resemble a mean-field description, where only a single coefficient would play a significant role. When the coefficient is distributed over significant number of indices, the corresponding many-body state is referred to a delocalized state. In the depicted Figs. (2a - 2d), representing $V_0$ values of $0.1E_r$, $0.3E_r$, $0.5E_r$, and $0.8E_r$, respectively, a considerable number of non-vanishing coefficients are evident. This suggests that the many-body state displays a degree of partial delocalization when the lattice depth is very small. With a further increase in lattice depth, Figs. (2e, 2f) the contributing coefficients spread out. Eventually, for the true Mott state depicted in Fig.~2(g) and (h), characterized by one-third fragmentation, the state becomes fully delocalized. Thus the fragmented SF (QSF) and incomplete fragmented Mott (QMI) exhibit partial delocalization in the Fock space whereas true Mott state exhibits complete delocalization. As the contributing coefficients spread out, indicating a delocalized state, predictions for QSF and QMI states go beyond mean-field calculations. Mean-field descriptions are insufficient to capture these phenomena.
It is crucial to emphasize that, although higher number of coefficients exhibit delocalization but they do not fully occupy the entirety of the allocated space designated for $M=12$. This particular aspect holds significant importance in guaranteeing the convergence of our results. 
\\
\begin{figure}
\centering
\includegraphics[scale=0.12, angle=-0]{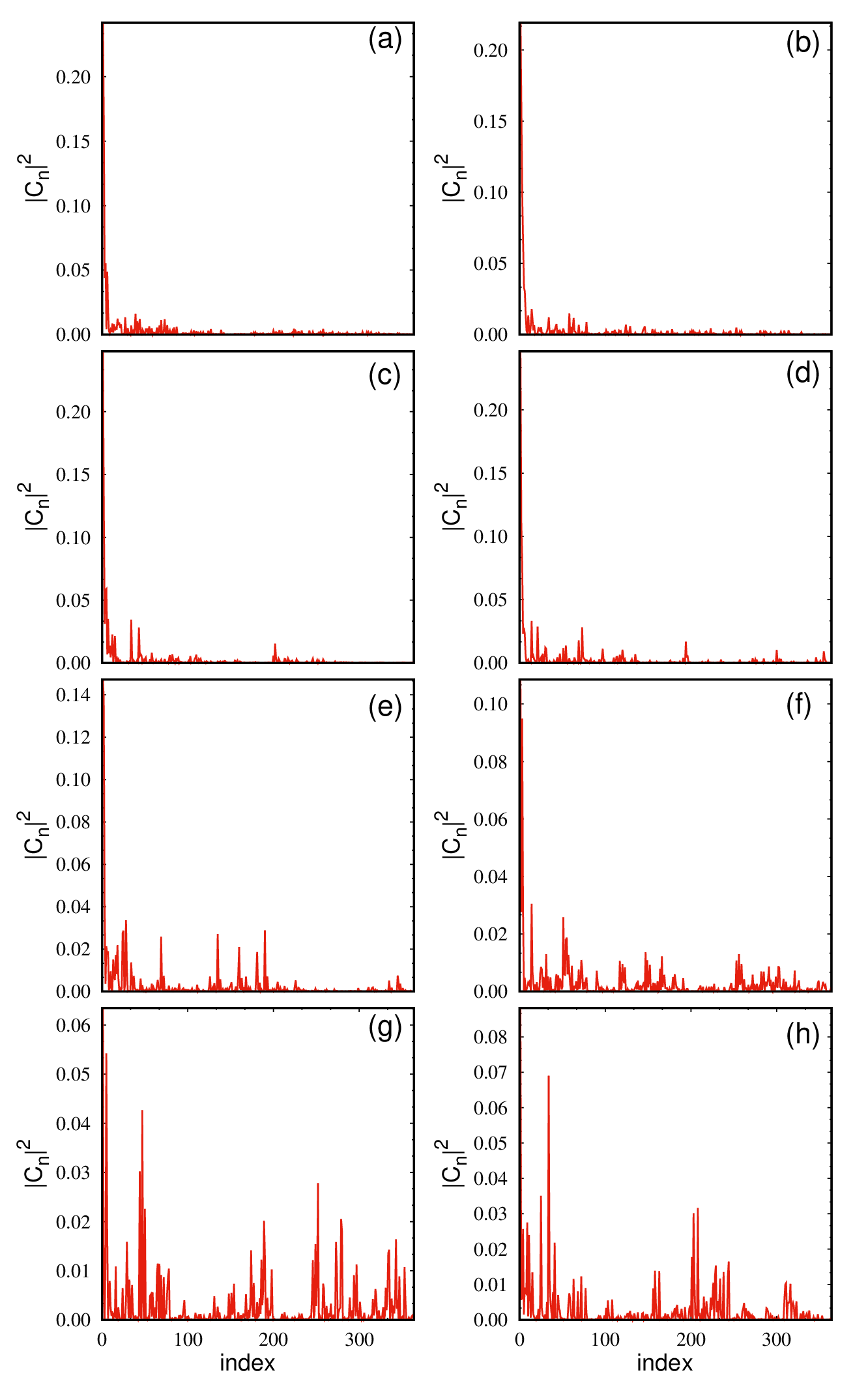}

\caption{Delocalization in Fock space in the fragmented SF (QSF), incomplete fragmented MI (QMI), and MI phases. The distribution of the magnitude of the coefficients is shown as a function of the index. Figures (a) to (h) correspond to $V_0$(in the unit of $E_r$) = $0.1, 0.3,0.5,0.8,1.0,2.0,5.0,8.0$ respectively and for fixed interaction strength $\lambda_0=10.0$. The transition from QSF to QMI to MI is exhibited by the transition from partial delocalization to complete delocalization in Fock space as determined by the $M=3$ orbitals. All quantities are dimensionless.}

\label{fig2}
\end{figure}

\subsection{One- and two-body correlation function}
The one-body and two-body Glauber's correlation functions are evaluated accordingly Eqs. (6) and (7)--they provide spatially resolved information about the coherence of atoms. 
The structure of $|g^{(1)}|^{2}$ with varying lattice depth gives an intuitive picture of the mechanism behind the excited fragmentation and also the underlying correlation in the complex many-body phases as observed in the vanishing and weak lattices. 
It also measures the proximity of the many-body state to a product of uncorrelated mean-field states for a fixed coordinate $(x, x^{\prime})$. 
Fig.~3 shows a plot of $|g^{(1)}(x, x^{\prime})|^{2}$ for various lattice depth. 
For $V_0=0.1E_r$, the uniform bright diagonal exhibits strong diagonal correlation only. 
It has no structure, and the isolated confinement of atoms in three distinct wells is not observed. 
Thus it is not a Mott phase although diagonal correlation is exhibited. 
The one-body correlation function also quantifies the off-diagonal correlation $(x \neq x^{\prime})$.
In the scale of diagonal correlation ($\simeq 100 \%$), the off-diagonal correlation measures $\simeq 40 \%$ to $\simeq 50 \%$ all throughout the lattice. 
It signifies that this is a very complex many-body phase intermediate to superfluid and Mott phase. 
The complexity is due to interplay between strong interaction and correlation, the effect of lattice is negligible. 
The strong repulsive interaction pushes the atoms out of the well-thus it lose diagonal structure and builds up the off-diagonal correlation. 
This is a fragmented superfluid phase and the structure of $|g^{(1)}|^{2}$ is in perfect agreement with the observations made in Fig.~1. 
In the same figure, we present results for various lattice depth $(V_0= 0.3E_r,0.5E_r, 0.8E_r, 1.0E_r, 2.0E_r, 5.0E_r, 8.0E_r)$.
One can follow how the structure in $|g^{(1)}|^{2}$ changes- as the lattice effect comes into play.
With increase in lattice depth, the diagonal correlation gradually develops structure and at the same time, the off-diagonal correlation is gradually reduced, which characterizes the incomplete fragmented Mott phase or QMI. 
With a further increase in lattice depth, the lattice effect becomes significant, and complete localization happens.
The three distinct lobes across the diagonal exhibits a clear signature of loss of inter-well correlation while maintaining intra-well correlation which characterizes the three-fold fragmented Mott phase. \\

\begin{figure}
\centering
\includegraphics[scale=0.2, angle=0]{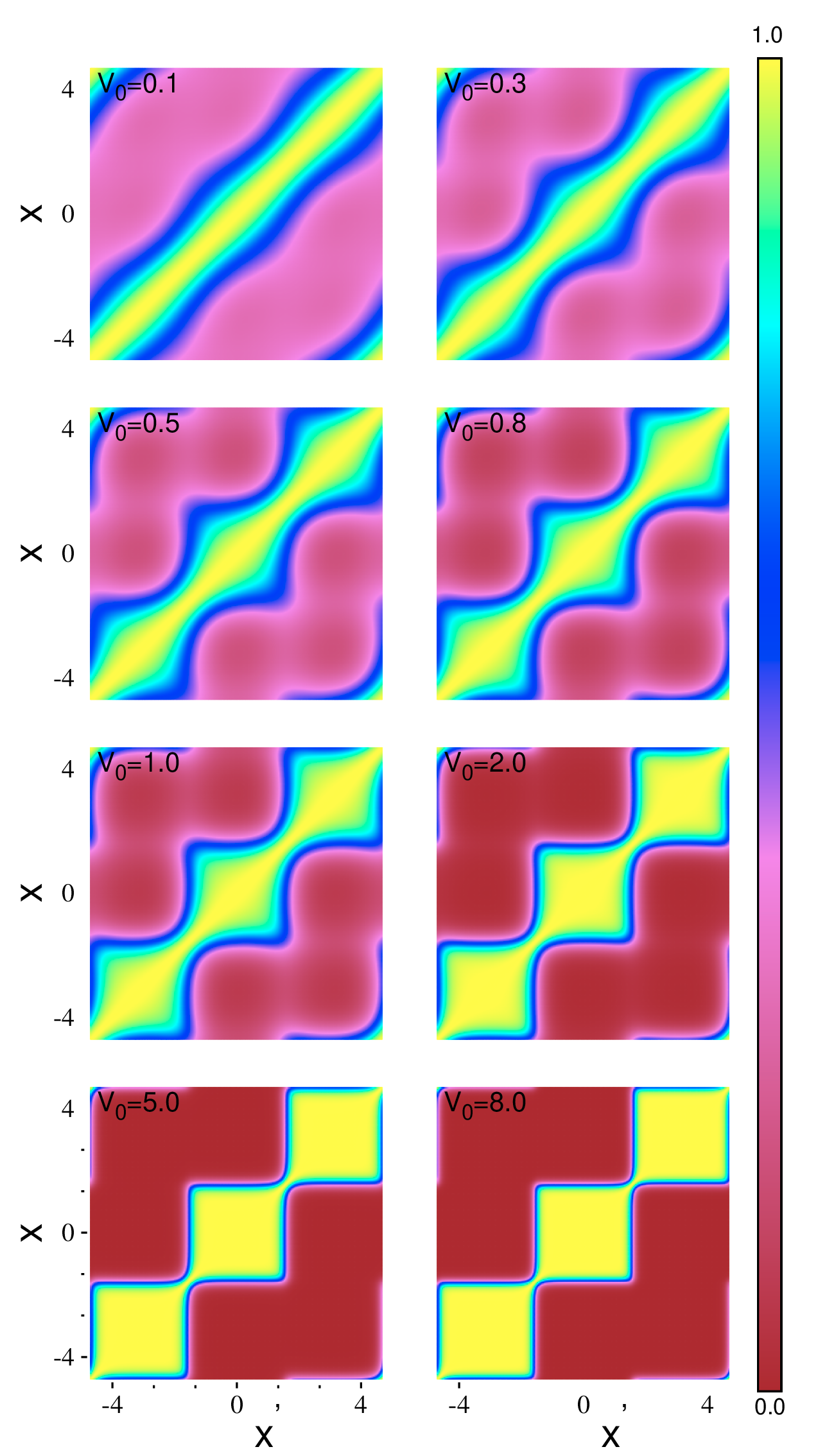}

\caption{First-order coherence of the fragmented SF (QSF), incomplete fragmented MI (QMI) and MI phases.
The normalized first-order correlation function $|g^{(1)} (x,x^{\prime})|^2$ is plotted for various lattice depth. For the fragmented SF phase in the vanishingly small lattice, structureless diagonal correlation, as well as off-diagonal correlation, are exhibited. The lattice effect is irrelevant in this phase. With the increase in the lattice depth,  the lattice effect and interatomic interaction compete and result in incomplete fragmented Mott (QMI). The structure in the diagonal line develops, but the off-diagonal correlation is not completely lost.  For $V_0= 5.0E_r$ and $V_0=8.0E_r$ the correlation function is characterized by three separated regions
with $|g^{(1)}|^{2} = 1$ along the diagonal and $|g^{(1)}|^{2} = 0$ on the off-diagonal.
Thus, the first-order coherence is maintained within wells and completely lost
between the wells; it characterizes the MI phase. All quantities are dimensionless.}

\label{fig4}
\end{figure}
%\end{document}

Next, we analyze the two-body correlation function $g^{(2)}(x^{\prime},x,x^{\prime},x)$ = $g^{(2)}(x,x^{\prime})$ for the same set of lattice depth potential as chosen earlier for the study of one-body correlation and plot it in Fig.~4. For a perfect superfluid phase (when bosons are weakly interacting) - $g^{(2)}(x,x^{\prime})$ $\simeq$ 1 for all $x$ and $x^{\prime}$--it infers that both the intra-well as well as inter-well coherence are maintained. For a Mott phase (deep lattice), $g^{(2)}$ $\simeq$ $1$ at the off-diagonal part and vanishes along the diagonal. However, strongly interacting bosons in very weak lattices exhibit interesting structures both along the diagonal and off-diagonal. For vanishingly small lattice depth  when the phase is 
fragmented superfluid, $g^{(2)}$ is extinguished along the diagonal; off-diagonal correlation is maintained across the lattice -- consistent with the observation in $g^{(1)}$. Now we gradually increase the lattice depth; $V_0$ =$0.3E_r, 0.5E_r, 0.8E_r$--the diagonal part develops a structure called ``correlation hole" or ``antibunching effect". Two particles can never be simultaneously found in the same well due to strong repulsion- this is termed as ``antibunching effect". The state is the incomplete fragmented Mott phase (QMI). Thus the fragmented superfluid and the incomplete fragmented MI both exhibit extinction of diagonal two-body correlation due to the antibunching effect, but the diagonal of the QSF phase is structureless whereas the QMI phase has the clear signature of correlation hole. With further increase in the lattice depth 
the system gradually approaches one-third fragmentation and when the fragmented Mott phase is 
reached,  the correlation hole becomes distinct and the three brown
lobes clearly distinguish three correlation holes.
\begin{figure}
\centering
\includegraphics[scale=0.2, angle=0]{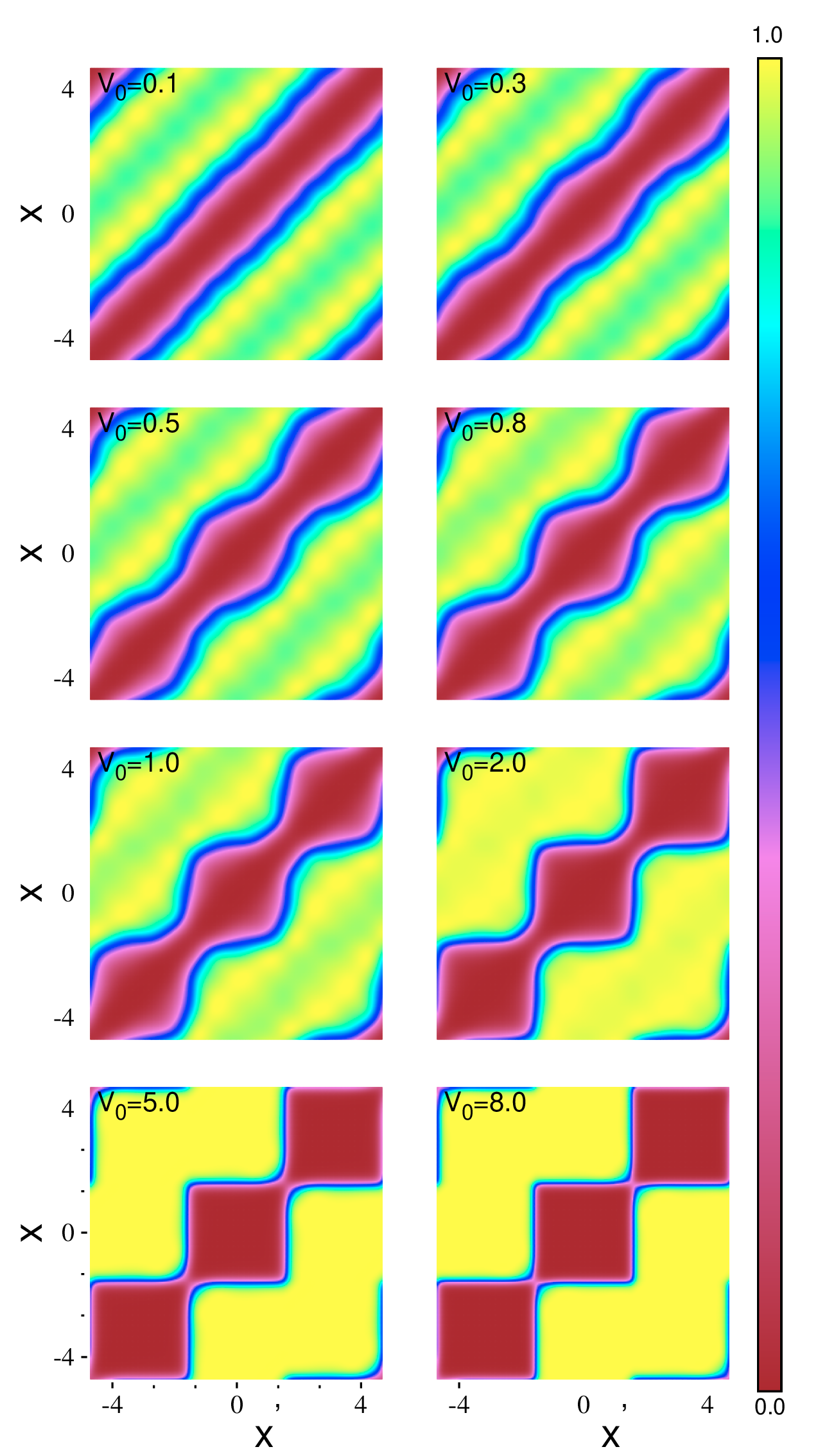}

\caption{Second-order coherence of fragmented SF (QSF), incomplete fragmented MI (QMI) and MI phases.
The normalized second-order correlation function $g^{(2)} (x,x^{\prime})$ is plotted for various lattice depth. A strong interplay between correlation and lattice effect is observed. 
A strong antibunching effect is demonstrated even for very weak lattices when the lattice effect is subrelevent. With the increase in lattice depth, gradually the structure of the correlation hole is built up and finally, it settles into three distinct correlation holes for the MI phase in the deep lattice.
All quantities are dimensionless. See text for further discussion.}

\label{fig5}
\end{figure}

\begin{figure}[ht!]
    \centering
    \includegraphics[width=0.4\textwidth]{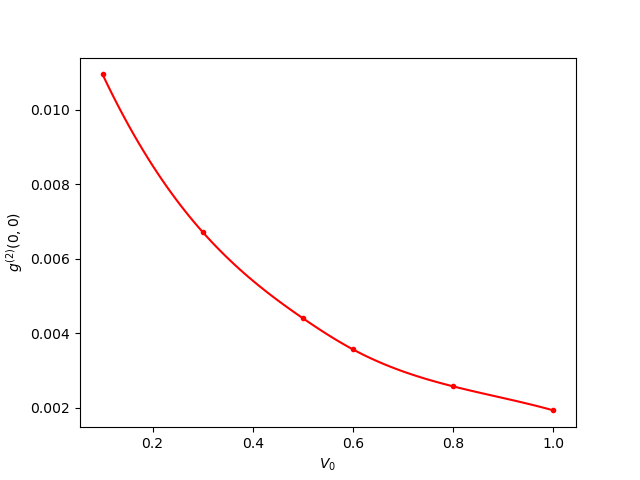}
    \caption{Variation of the local two-body correlation function, $g^{(2)}(0,0)$, with lattice depth in the shallow optical depth regime. As the lattice depth increases, $g^{(2)}(0,0)$ exhibits a gradual decrease. Further details are provided in the text. The quantity is dimensionless.}
    \label{fig6}
\end{figure}
Fig.~\ref{fig6} shows the local two-body density-density correlation $g^{(2)}(0,0)$ as a function of lattice depth only in the regime of vanishing small and weak lattices. 
This image illustrates the relationship between lattice depth and interaction strength. In the case of significant repulsive interactions, the bosons exhibit a tendency to repel each other, leading to a near-zero value for the second-order correlation function at $x=0$ and $x^{\prime} = 0$. A similar outcome is observed in the Mott Insulator phase, where each boson occupies a distinct well, resulting in an ideal Mott-Insulator state with $g^{(2)}(0,0)$ precisely equals to zero.
Throughout the range of lattice depths, we consistently observe a low value for the second-order local correlation function, $g^{(2)}(0,0)$. This indicates that the bosons, which interact strongly, tend to avoid each other. As we increase the lattice depth, $g^{(2)}(0,0)$ gradually decreases.  
 The change in $g^{(2)}(0,0)$ maps the system's journey through different phases. Beginning with a small $V_0$, where $g^{(2)}(0,0)$ is also small, it indicates a quasi-superfluid state in a vanishing lattice with strong interactions. The gradual decrease in $g^{(2)}(0,0)$ marks a shift from a quasi-SF to quasi-Mott insulator (QMI) phase where $g^{(2)}(0,0)$ is not precisely zero. As we move to larger lattice depths, the system enters the Mott insulator (MI) phase, where $g^{(2)}(0,0)$ becomes exactly zero (not depicted in the figure).\\
 
\begin{figure}[ht!]
    \centering
    \includegraphics[width=0.4\textwidth]{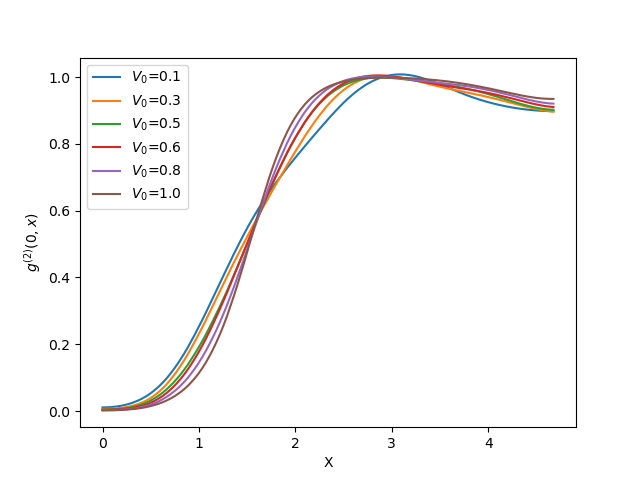}
    \caption{Nonlocal density-density correlation at various lattice depths in the shallow lattice depth regime. Initially, $g^{(2)}(0,x)$ exhibits a low value near zero, gradually increases, and peaks around $x\sim 3.14$. This figure illustrates the persistence of two-body correlation with the neighboring lattice site and loss of two-body correlation within the lattice. The quantity is dimensionless. Additional details can be found in the text.}
    \label{fig7}
\end{figure}

The nonlocal density-density correlation $g^{(2)}(0,x)$ is plotted in Fig.~\ref{fig7}.
Due to a special symmetry in position space on either side of $x=0$, we opt to plot only one half ($x>0$) for various lattice depths.
At $x=x^\prime=0$, the local correlation remains consistently low, approaching zero due to strong repulsive interaction; the value is smaller for larger lattice depth as expected.
The second-order non local correlation function $g^{(2)}(0,x)$ consistently increases and peaks around $x\approx 3.14$. This is significant because in a lattice with periodicity $\pi$, the minimum of the next lattice site aligns with $x\approx 3.14$. Considering three lattice sites, $g^{(2)}(0,x)$ exhibits a high value around $x\approx 3.14$, indicating strong second-order correlations between adjacent sites. Moreover, the values of $g^{(2)}(0,x\approx 3.14)$ are higher for larger $V_0$. This highlights the combined influence of repulsive interaction and lattice depth in weak optical lattice. Thus, $g^{(2)}(0,x)$ serves as a valuable metric for studying the transition from quasi-superfluid (QSF) to quasi-Mott insulator (QMI) phase. 

\subsection{Measures of information entropy}

Many-body information entropy is another useful quantity for further characterization of complex many-body phases. We exemplify the entropy measures with the fragmentation and delocalization. We calculate the many-body information entropy $S^{info}$ ($S_c$) using the expansion coefficient (Eq.(8)) and occupation entropy $S^{occu}$ ($S_n$) by the eigenvalues of the reduced one-body density matrix (Eq.(9)). We plot $S^{info}$ and $S^{occu}$ as a function of lattice depth in Fig.~7. In the mean-field description as a single coefficient contributes, information entropy is always zero. In the many-body picture, the behavior of $S_c$ is controlled by the way the many-body wave function spreads over the Fock space. According to Figures (Fig.~1 and Fig.~2), as the many-body state is initially fragmented and delocalized, the initial entropy is significant. With the increase in lattice depth as the Fock space is populated more, $S_c$ increases. When Fock space is filled up, all coefficients are large, $S_c$ saturates. \\
As occupation entropy is obtained from the natural occupation, it is always zero for the mean-field picture, when there is only one natural occupation. We find, for a very weak lattice, when the many-body state is already fragmented, several natural orbitals contribute, and $S_n$ has a significant value. With the increase in lattice depth, the many-body state gradually approaches one-third fragmentation--$S_n$ saturates.\\
We make a comparison between the saturation values of entropy and the prediction of the Gaussian orthogonal ensemble of random matrices (GOE)~\cite{Kota}. For GOE random matrices $S_{c}^{GOE}= \ln(0.48D)$, where $D \times D$ is the dimension of the random matrices. In our calculation, $D$ is equal to the number of contributing many-body states which are close to $340$ (as seen in Fig. 2(h)), thus  $S_{c}^{GOE} = 5.09$, whereas our numerical value of saturation is $4.61$ (Fig. 7(a)). For the occupation entropy, the GOE analog is obtained by setting $n_i=\frac{N}{M}$, and consequently $D$ becomes the number of orbitals $M$. Thus we obtain $S_{n}^{GOE}= \ln(M)$. For a three-fold fragmented state, the lowest three natural orbitals contribute, thus $M=3$ and $S_{n}^{GOE} =1.09$ which is the saturation value as shown in Fig. 7(b).  \\

\begin{figure}
\centering
\includegraphics[scale=0.22, angle=-0]{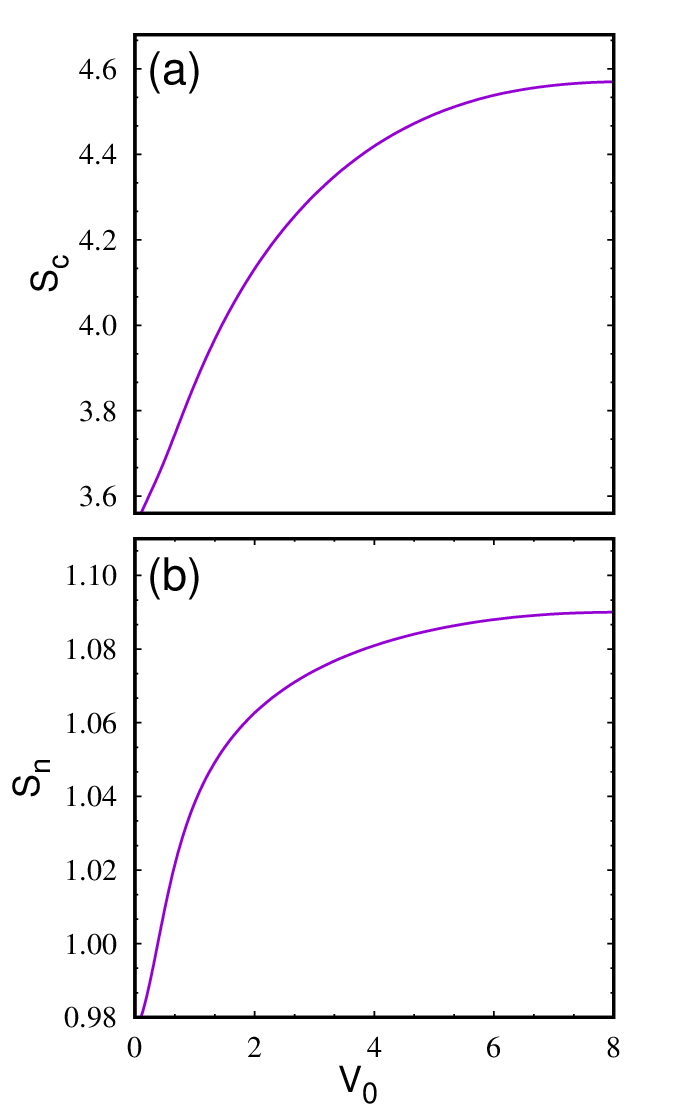}

\caption{Plot of many-body entropy as a function of lattice depth $V_0$. (a) corresponds to information entropy $S_c$ and (b) corresponds to occupation entropy $S_n$. Initially, the system has significant entropy as the many-body state is fragmented and delocalized. Both measures saturate at large lattice depth when the system makes a transition to a three-fold fragmented, completely delocalized Mott phase. All quantities are dimensionless. See the text for additional discussion.}

\label{fig3}
\end{figure}

\section{Conclusion}
In conclusion, we have applied a numerically exact many-body approach, MCTDHB, to solve the many-body Schr\"odinger equation to study the ground states of strongly interacting bosons in vanishingly small and weak lattices. We explore the emergent phases: fragmented superfluid and incomplete fragmented Mott phases which are termed quasi superfluid (QSF) and quasi Mott (QMI) phases. In the usual notion of the Bose Hubbard picture: the superfluid is characterized by a single orbital occupation that exhibits correlation across the lattice; whereas the Mott phase is characterized by complete localization in each site and thus extinction of inter-well correlation. However, in the many-body perspective, we find some intermediate phases that can not be explored by mean-field theory or the Bose-Hubbard model. To differentiate between these recently discovered phases, we have utilized fragmentation as a key metric. Additionally, we have introduced an order parameter based on the eigenvalues of the reduced one-body density matrix, providing another effective tool for investigating these distinct phases. Our analysis has been further bolstered by comprehensive study of one- and two-body correlations, as well as information entropy.
Fragmentation is the figure of merit to distinguish the fragmented superfluid phase from the traditional superfluid phase as well as the incomplete fragmented Mott phase from the traditional Mott phase. We study the complete pathway of transition to the Mott phase for deep lattices. The measures of several information entropy clearly distinguish the limitation of mean-field physics. The order parameter and the structure of one and two-body coherence are utilized as the key markers to study the new phases. The structures evident in one- and two-body coherence serve as robust indicators, facilitating an understanding different intermediate phases in optical lattice.
We work with a finite ensemble of few bosons but this provides an experimental bottom-up access to the many-body physics. The dynamic evolution from the superfluid (SF) phase to the Mott insulator (MI) phase has already been explored~\cite{rhombik_quantumreports,rhombik_jpb,PhysRevB.40.546}. An immediate and open question in this line is to investigate how the system evolves when subjected to a quench from the QSF to the QMI phase in the strong interaction regime. It would also be a challenging many-body calculation to study the stability of these intermediate phases in the presence of disorder.

\section*{Acknowledgements}
A G acknowledges CNPq-Conselho Nacional de Desenvolvimento Científico e Tecnológico (Brazil) grant no. 306209/2022-0. B C acknowledges FAPESP grant Process No. 2023/06550-4.

%\bibliographystyle{elsarticle-harv} 
%\bibliography{ref}



%% For example:

%% \bibitem[Aladro et al.(2015)]{Aladro15} Aladro, R., Martín, S., Riquelme, D., et al. 2015, \aas, 579, A101

%%\end{thebibliography}

\end{document}